\documentstyle[prb,aps,floats,epsf,multicol,graphics]{revtex}

\newcommand{\smeq}{\! = \!}

\newcommand{\smpl}{\! + \!}
\newcommand{\smmi}{\! - \!}
\def\lsim{\hbox{\lower .8ex\hbox{$\, \buildrel < \over \sim\,$}}}
\def\gsim{\hbox{\lower .8ex\hbox{$\, \buildrel > \over \sim\,$}}}

\newcommand{\e}{\epsilon}
\newcommand{\ve}{\varepsilon}

\newcommand{\Ef}{E_{F}}
\newcommand{\Eth}{E_{\rm th}}

\newcommand{\ag}{{\cal N}}
\newcommand{\kt}{k_{ B}T}
\newcommand{\Egs}{E_{\scriptscriptstyle GS}}
\newcommand{\be}{\begin{equation}}
\newcommand{\ee}{\end{equation}}
\newcommand{\Ha}{{\hat H}}

\epsfclipon

\begin{document}
\twocolumn[\hsize\textwidth\columnwidth\hsize\csname
@twocolumnfalse\endcsname
\draft
\title{Coulomb Blockade Peak Spacing Distribution: The Interplay of Temperature and Spin}
\author{Gonzalo Usaj and Harold U. Baranger}
\address{Department of Physics, Duke University, PoBox 90305, Durham NC 27708-0305}
\date{\today}
\maketitle 

    \begin{abstract}
We calculate the Coulomb Blockade peak spacing distribution at finite
temperature using the recently introduced ``universal Hamiltonian'' 
to describe the $e$-$e$ interactions. We show that the temperature effect is
important even at $\kt\!\sim\!0.1\Delta$ ($\Delta$ is the single-particle
mean level spacing). This sensitivity arises because: (1) exchange
reduces the minimum energy of excitation from the ground state and
(2) the entropic contribution depends on the change of the spin of the
quantum dot. Including the leading corrections to the universal Hamiltonian 
yields results in quantitative agreement with the experiments.
Surprisingly, temperature appears to be the most important effect.
    \end{abstract}
]  
\narrowtext
\bigskip

Among the unique features of quantum dots (QDs) is the possibility to control their number of electrons 
$N$.~\cite{KouwenetalRev97} This is done by weakly coupling a QD to source and drain leads and using a 
gate voltage $V_g$ to control its electrostatic potential. When the thermal energy $\kt$ is smaller 
than the charging energy $E_{C}$ required to add an electron to the QD, the electron transport is blocked 
and $N$ is fixed. By sweeping 
$V_g^{}$ this Coulomb Blockade (CB) effect can be overcome at the particular value $V_g^{N}$ where the transition
$N\!\rightarrow\!N+1$ occurs. The conductance $G(V_{g}^{})$ shows then a series of sharp peaks as a 
function of $V_g^{}$.

At sufficiently low temperature, $\kt\!\ll\!\delta$ where $\delta$ is the energy gap between the ground state (GS) and the first
 excited state of the QD, only the former contributes significantly to the conductance peak. In that case, the position of the CB 
peak is proportional to the change of the GS energy of the QD upon the addition of one electron.
\cite{KouwenetalRev97} Therefore, the CB peak spacing distribution (PSD) yields information about the 
{\em many-body} GS properties of the QD.

This has been the subject of experimental
\cite{SivanBAPAB96,PatelCSHMDHCG98,PatelSMGASDH98,SimmelAWKK99,LuscherHEWB01} 
and theoretical
\cite{PrusAASB96,BlanterMM97,Berkovits98,BrouwerOH99,BarangerUG00,KurlandAA00,JacquodS00,UllmoB01,VallejosLM98}
 work over the last years. One reason is that the simplest model used for the CB conductance peaks fails 
drastically in describing the observed PSD. It assumes a constant $e$-$e$ interaction ($E_C$)---hence the name constant 
interaction (CI) model.
Recently, however, it has become clear that residual $e$-$e$ interactions (i.e., those beyond $E_C$) 
play an important role in determining the GS of the QD
\cite{PrusAASB96,BlanterMM97,Berkovits98,BrouwerOH99,BarangerUG00,KurlandAA00,JacquodS00,UllmoB01} 
and therefore must be included in the description of the PSD.
In particular, the addition of the average exchange interaction,
\cite{BrouwerOH99,BarangerUG00,KurlandAA00} which gives the ``universal Hamiltonian'', hereafter called the constant 
exchange and interaction (CEI) model,  leads to a {\em completely} different PSD.\cite{UllmoB01}
Yet, this is not enough to account for the observed distribution: other (smaller) contributions---such as the 
``scrambling" effect,\cite{BlanterMM97} ``gate" perturbations\cite{VallejosLM98} and the fluctuation of the 
interactions\cite{UllmoB01}---have to be considered.\cite{UsajB01_universal} So far, however, a much simpler effect has 
not been considered within the CEI model: the effect of finite temperature.

The goal of this work is twofold. First, we show that in the CEI model the temperature effects are more important than in the CI 
model and that they become significant even at $\kt\!\sim\!0.1\Delta$ where $\Delta$ is the single-particle mean level spacing. 
In particular, the {\em shape} of the peak spacing distribution changes significantly while increasing temperature. Since most 
experiments were done in the regime $\kt\!\sim\!0.3$-$0.5\Delta$---an exception is Ref.\onlinecite{LuscherHEWB01}---our  
results are crucial for interpreting the experimental data. 
Second, we calculate the PSD including {\em all} the leading order 
corrections to the CEI model. The final result for the distribution is in {\em quantitative} agreement with the
experimental data of
 Refs.~\onlinecite{PatelCSHMDHCG98} and \onlinecite{OngBHPM01} once the temperature effect  is included. 
Surprisingly, the latter introduces 
the biggest correction to the $T\smeq0$ CEI model result. 

Mesoscopic fluctuations associated with single-particle properties of chaotic QDs are known to be well described by random 
matrix theory (RMT) in an energy window up to the Thouless energy $\Eth$. The treatment of the $e$-$e$ interaction is more subtle 
however. From Fermi-liquid theory, we expect the screening of the Coulomb interaction to be important for $N\!\gg\!1$. 
In that case, the residual interaction should be weak, and perturbative treatments, such as RPA, seem adequate. Using such an 
approach and RMT to describe the single-particle Hamiltonian, it is possible to derive\cite{KurlandAA00,AleinerBG01} an 
effective Hamiltonian $\Ha_{\mathrm QD}$ for the QD.  The small parameter in the perturbation theory is 
$g^{\smmi1}\!\propto\!N^{\smmi\frac{1}{2}}\!\ll\!1$ with $g\smeq\Eth/\Delta$ the dimensionless conductance. The zeroth-order term 
($g\!\rightarrow\!\infty$) in this expansion corresponds to the ``universal Hamiltonian'' and is given by
\cite{KurlandAA00,AleinerBG01} 
\be
\Ha_{\mathrm QD}^{(0)}\smeq\sum_{\alpha,\sigma}{\ve_{\alpha}}\,\hat{n}_{\alpha,\sigma}
                              \smpl E_{C}\,(\hat{n}-\ag)^2\smmi J_{S}\;{\vec S}^2
\label{uH}
\ee
where $\{\ve_{\alpha}\}$ are the single-electron energies, $\ag\smeq C_g V_g/e$ describes the capacitive coupling to the 
control gate, ${\vec S}$
is the total spin operator, and  $J_{S}$ is the exchange constant. The difference between the CEI and 
CI models is the additional term proportional to ${\vec S}^2$. Since $J_S\!<\!\Delta/2$ has a fixed value, the mesoscopic 
fluctuations in the spectrum of $\Ha_{\mathrm QD}^{(0)}$ arise only from $\{\ve_{\alpha}\}$.
This is a key point for understanding its GS: while in the CI model the levels are filled in an ``up-down" scheme---which 
leads to a bimodal PSD---in the CEI model it is energetically favorable to promote an electron to a higher level and gain exchange 
energy whenever the spacing between two consecutive single-particle levels is smaller than $2 J_S$ (for $N$ even). This leads to a 
GS with $S\!\ge\!1$ and the simple ``up-down" filling scheme breaks down.\cite{BrouwerOH99,BarangerUG00,KurlandAA00} 
Consequently, the PSD is very different\cite{UllmoB01} from the CI model result (see Fig.~1).

\begin{figure}[t]
\begin{center}
\leavevmode
\rotatebox{-0}{
\epsfxsize = 7cm 
\epsffile{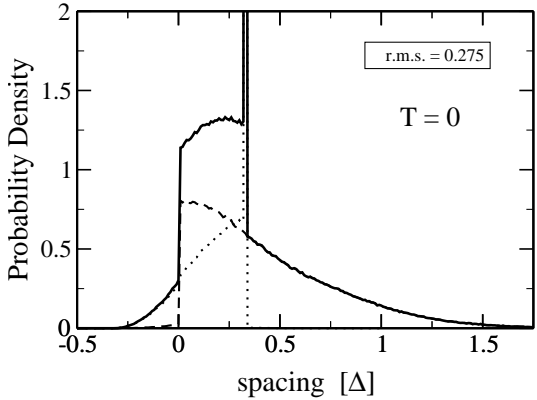}
}
\end{center}
\caption{CB peak spacing distribution obtained using the CEI model (Eq. \ref{uH}) with $T\smeq0$ and $J_S\smeq0.32\Delta$ 
($r_s\!\approx\!1.5$). The dashed (dotted) line shows the $N$  even (odd) contribution (each spacing corresponds to a 
$N\smmi1\!\rightarrow\!N\!\rightarrow\!N\smpl1$ transition). Notice that the strong even/odd effect predicted by the CI 
model is significantly reduced and that the $\delta$-function in the odd distribution persists---but is shifted by $J_S$. 
The zero in the horizontal axis corresponds to $2 E_C$ in the CI model.} 
\end{figure}
The $T\smeq0$ distribution shown in Fig.~1 will remain a good description so long as the contribution from the 
excited states can be ignored. To estimate that, let us calculate the average occupation ${\cal P}$ of the first excited state  
assuming that only it and the GS are relevant. In the CI model, the energy gap $\delta$ 
between those states is $\delta\smeq\Delta\ve$ where $\Delta\ve$ is the single-particle energy spacing 
 between the top levels. Using the GUE Wigner-Dyson distribution for $\Delta\ve$---broken time-reversal symmetry is
 assumed throughout---we find ${\cal P}\smeq0.02$ for $\kt\smeq0.1\Delta$. Thus,
the excited states can indeed be ignored for $\kt\!\le\!0.1\Delta$.  This is {\em not} the case in the CEI 
model, where $\delta$ gets reduced by the exchange interaction. For instance, for $N$ even, 
$\delta\smeq|\Delta\ve\smmi2J_S|$ and  ${\cal P}\smeq0.14$ with $J_S\smeq0.32\Delta$. 
Therefore, exchange not only modifies the $T\smeq0$ distribution but makes the temperature effect stronger.  

Furthermore, at finite temperature, the peak
position involves the change in {\em free energy} of the QD upon adding a particle. 
Then, the spin degeneracy should play an important role through the entropy contribution.
\cite{GlazmanM88,Beenakker91,Akera99a}
Let us consider the regime $\Gamma\ll \kt,\Delta\ll E_C$, where 
$\Gamma$ is the total with of a level in the QD. Near the CB peak corresponding to the 
$N\smmi1\!\rightarrow\!N$ transition, the linear conductance is given by
\cite{Beenakker91,MeirWL91} 
\be
G(\ag)\smeq\frac{e^2}{\hbar \kt}\, P_{\mathrm eq}^{N}\,
                \sum_{\alpha}{\frac{\Gamma_{\alpha}^L\Gamma_{\alpha}^R}
                 {\Gamma_{\alpha}^L\smpl\Gamma_{\alpha}^R}\,w_{\alpha}}
\label{G}
\ee
with $\Gamma^{L(R)}_{\alpha}$ the partial width of the single-particle level $\alpha$ due to tunneling to the left (right) lead and
\be
w_{\alpha}\smeq\sum_{i,j,\sigma}{F_{\mathrm eq}(j|N)
         \left|\langle\Psi_{j}^{N}|c_{\alpha,\sigma}^{\dagger}|\Psi_{i}^{N\smmi1}\rangle\right|^2
              [1\smmi f(\e_{j}\smmi\e_{i})]}.
\label{weight}
\ee
Here $P_{\mathrm eq}^{N}$ is the {\em equilibrium} probability that the QD contains $N$ electrons, 
$\Ha_{\mathrm QD}|\Psi_{j}^N\rangle\smeq\e_{j}|\Psi_{j}^N\rangle$, 
$F_{\mathrm eq}(j|N)$ is the conditional probability that the eigenstate $j$ is occupied given 
that the QD contains $N$ electrons, and $f(\e)\smeq\{1\!+\exp[(\e-\!\Ef)/\kt]\}^{-1}$. 
Since near the peak only the states with $N\smmi1$ and $N$ electrons are relevant, we have 
$P_{\mathrm eq}^{N}\!\simeq\!f({\cal F}_{N}\smmi{\cal F}_{N\smmi1})$ with ${\cal F}_N$ the canonical free energy of the QD.
\cite{Beenakker91}
To make the dependence on $\ag$ explicit, let us denote by $\{E_{j}^{}\}$ the eigenenergies of 
$\Ha_{\mathrm QD}$ without the charging energy term. Then, 
$\e_{j}\smmi\e_{i}\smeq E_{j}^{N}\smmi E_{i}^{N\smmi1}\smpl 2E_C \delta\ag$ and 
${\cal F}_{N}\smmi{\cal F}_{N\smmi1}\smeq\mu_{N}\smpl 2E_C \delta\ag$ with  $\delta\ag\smeq(N\smmi\frac{1}{2})\smmi\ag$ and 
$\mu_{N}\smeq E_{j}^{N}\smmi E_{i}^{N\smmi1}\smpl\kt\ln\!\left[F_{\mathrm eq}(j|N)/F_{\mathrm eq}(i|N\smmi1)\right]$
for any $i$ and $j$. 
The contribution of the transition $i\!\rightarrow\!j$ to the conductance reaches its maximum when 
\be
\Ef\smeq E_{j}^{N}\smmi E_{i}^{N\smmi1}
\smpl\frac{\kt}{2}\ln\left[\frac{F_{\mathrm eq}(j|N)}{F_{\mathrm eq}(i|N\smmi1)}\right]
\smpl2E_C\delta\ag.
\label{gcond} 
\ee

In the particular case where the  transition between GS dominates, and taking the spin degeneracy into account, the CB peak 
position is given by
\be 
\Ef\smeq\Egs^{N}\smmi\Egs^{N\smmi1}\smmi\frac{1}{2}
\kt\ln\left[\frac{2S_{\scriptscriptstyle GS}^N\smpl1}{2S_{\scriptscriptstyle GS}^{N\smmi1}\smpl 1}\right]\smpl2E_C\delta\ag.
\label{cond}
\ee
We see that the peak is {\em shifted} with respect to its position at 
$T\smeq0$ by an amount depending on the change of the spin of the QD.\cite{GlazmanM88,Beenakker91,Akera99a} 
Because the r.m.s. of the PSD is $\!\sim\!0.3\Delta$ (see Fig.~1), this shift is 
significant {\em even} for $\kt\!\sim\!0.1\Delta$. 
In addition, while in the CI model this introduces only a constant shift between the even 
and odd distributions, in the CEI model it changes the {\em shape} of both distributions 
since different spin transitions contribute to each one.

Note that because of this shift, the on-peak conductance is 
renormalized.\cite{GlazmanM88,Beenakker91} Since different spin transitions lead to different renormalizations, the average 
conductance peak depends not only on the average coupling to the leads but also on $J_S$ and on the
statistics of the spectrum. 
This explains the small deviations observed\cite{FolkMH00} at low temperature from the values predicted in the absence of the 
exchange interaction (in particular the increase in Fig.~2 of Ref. \onlinecite{FolkMH00}). 

In the general case more than one transition contributes to the conductance, and the 
CB peak position must be determined by maximizing Eq.(\ref{G}) with respect to $\ag$. For simplicity, we restrict 
ourselves to the case where only the GS and first two excited states are relevant. By comparing the peak 
spacing distribution with and without including the second excited state, we found that this is the case 
for $\kt\le0.2\Delta$, which is close to the experimental regime. 

\begin{figure}[t]
\begin{center}
\leavevmode
\rotatebox{-0}{
\epsfxsize = 7cm 
\epsffile{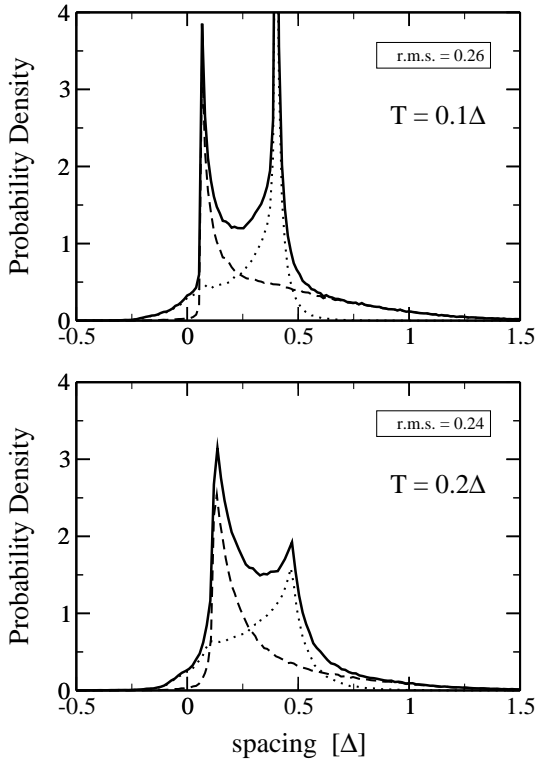}
}
\end{center}
\caption{Finite temperature CB peak spacing distribution obtained using the CEI model [Eq.~(\ref{uH})]. Here 
$J_S \smeq 0.32 \Delta$ and $\kt \smeq 0.1 \Delta$ $(0.2 \Delta)$ in the top (bottom) plot.
Notice that the $\delta$-function is 
smeared out and shifted by $\kt\ln2$ and that the even distribution develops a peak. 
}
\end{figure}

Figure 2 shows the PSD for non-zero temperature. The value $J_S\smeq0.32\Delta$ corresponds to a gas parameter 
$r_s \!\sim\! 1.5$. 
As expected, the sharp features are smeared out by temperature, including the $\delta$-function in the odd distribution. But 
more important, they are {\em shifted} due to the entropic term in (\ref{cond}). This is particularly clear for the peak associated 
with the $\delta$-function: it occurs at $J_S$ in Fig. 1 but at $J_S\smpl\kt\ln2$ in Fig. 2 (the $\delta$-function is always 
associated with the spin transition $0\!\rightarrow\frac{1}{2}\!\rightarrow0$). 
There are two other important effects worth comment: a) the even distribution {\em develops} a peak at 
small spacings while the odd one gets broader---in particular, at $\kt\smeq0.2\Delta$ the maximum of the total 
distribution is dominated by the even distribution, in contrast to what occurs at $T\smeq0$; b) the relative weight of the 
long tail of the even distribution is strongly reduced and the distribution becomes more symmetric.  

The peak in the even distribution arises from cases where $S\smeq1$ and $S\smeq0$ states are  (almost) degenerate---it 
corresponds to the sharp jump at zero spacing in Fig.~1. According to Eq.~(\ref{gcond}) the CB peak is shifted by $\frac{1}{2}\kt\ln(4/2)$ 
in that case, which gives a shift of $\kt\ln2$ for the peak in the PSD.
The deviation from the $T\smeq0$ result is still noticeable for 
$\kt\sim0.05\Delta$ (data not shown) which is the temperature in Ref.\onlinecite{LuscherHEWB01}.
We found that the r.m.s. of the total distribution decreases monotonically when increasing $T$ in the range $\kt\le0.2\Delta$, 
in contrast to the results obtained for the CI model.\cite{AlhassidM99} The available experimental 
data\cite{PatelSMGASDH98} cannot discern this difference.
An expected effect of the temperature is to increase the probability that $S\smeq1$. This can be 
easily estimated by calculating the average probability $\langle F_{\mathrm eq}(S\smeq1|N)\rangle$. We find
$\langle F_{\mathrm eq}(S\smeq1|N)\rangle\smeq0.21 (0.31)$ for $\kt\smeq0\ (0.1\Delta)$.

So far we have considered only the mean value of the residual interactions. It is clear that even at this level of approximation, 
the finite temperature PSD is quite different from the widely used $T\smeq0$ CI model result: {\em we should expect only a weak even/odd 
effect or asymmetry in the experimental data for $\kt\!\gsim\!0.2\Delta$}.
 
Nevertheless, the distribution still does not agree with the one observed experimentally. Thus, we must go to the next level 
of approximation and include the leading order corrections to $\Ha_{\mathrm QD}$. 
There are three contributions that lead to corrections of order $\Delta/\sqrt{ g}$ to the spacing: 
1) the ``scrambling" of the spectrum when adding an electron to the QD that originates in the 
change of the charge distribution in the QD;\cite{BlanterMM97} 
2) the change in the single-electron energies when the gate voltage is swept;\cite{VallejosLM98} 
3) the fluctuation of the diagonal part of the $e$-$e$ interaction.\cite{UllmoB01} 
After including them, the Hamiltonian of the QD reads\cite{AleinerBG01} 
\begin{eqnarray}
\Ha_{\mathrm QD}&\smeq& \Ha_{\mathrm QD}^{(0)}
               \smpl\frac{1}{2}
               \sum_{\alpha,\beta,\gamma,\delta}{H_{\alpha,\beta,\gamma,\delta}^{(1/g)}\;
               \; c^{\dagger}_{\delta,\sigma}c^{\dagger}_{\gamma,\sigma'}
               c^{}_{\beta,\sigma'}c^{}_{\alpha,\sigma}}
\nonumber \\
        &     & 
               \smpl\sum_{\alpha,\beta,\sigma}\,{c^{\dagger}_{\alpha,\sigma}c^{}_{\beta,\sigma}
                 \,[(\hat{n}-\ag){\cal X}^{0}_{\alpha,\beta}
                 \smpl\delta\ag\,{\cal X}^{1}_{\alpha,\beta}}]
\label{leading}
\end{eqnarray}
where, because of the fluctuations of the single-electron wavefunctions, the matrix elements 
$H_{\alpha,\beta,\gamma,\delta}^{(1/g)}$ and ${\cal X}^{j}_{\alpha,\beta}$ are gaussian random variables. They are characterized by 
${\mathrm var}({\cal X}^{j}_{\alpha,\beta})\smeq b_{jj}\Delta^{2}/g$ and 
${\mathrm var}(H_{\alpha,\beta,\gamma,\delta}^{(1/g)})\smeq c_{2}\Delta^{2}\ln({\tilde c_2}g)/g^{2}$; their mean values can be 
included in the definition of $E_C$ and $J_S$, so that 
$\langle H_{\alpha,\beta,\gamma,\delta}^{(1/g)}\rangle\smeq\langle{\cal X}^{j}_{\alpha,\beta}\rangle\smeq0$. 
Only the diagonal terms of the residual interaction (i.e. those where the operators 
$c^\dagger$ and $c$ are paired) are included in Eq.~(\ref{leading})---the off-diagonal terms can be neglected when calculating the 
PSD.\cite{JacquodS00,UsajB01_universal}
Here, $\delta\ag$ is taken with respect to the state with $N$ electrons. 
We use $g\smeq0.384\sqrt{N/2}$ for the dimensionless conductance, which 
corresponds to a disc geometry.\cite{AleinerBG01,BlanterMM97} 
The coefficients $b_{jj}$, $c_2$ and ${\tilde c_2}$ are geometry-dependent. We estimate\cite{UsajB01_universal} 
$b_{00}\!\simeq\!0.01$,\cite{comment} $c_2\!\simeq\!0.003$, ${\tilde c_2}\!\simeq\!37$ for the experiment in Ref. 
\onlinecite{PatelCSHMDHCG98}, and we 
assume $b_{11}\smeq b_{00}$. 
Notice that the magnitude of the ``scrambling" and ``gate" effects [last term in Eq.~(\ref{leading})] are much smaller than used 
previously in the literature.\cite{VallejosLM98,AlhassidM99} 

Figure 3 shows the PSD including these corrections to the CEI model.  
The additional fluctuations increase the smearing of the remaining pronounced features of the distribution as well as of its r.m.s.. 
The distribution is less asymmetric but the even/odd effect is still noticeable---it should be kept in mind that the experimental 
noise will contribute significantly to this smearing.  At $\kt\smeq0.2\Delta$ the peak of the 
distribution is still dominated by the even distribution. It is important to emphasize that this particular feature is exclusively related 
to the temperature effect.
A detailed analysis of the experimental data of Ref. \onlinecite{PatelCSHMDHCG98} 
shows\cite{OngBHPM01} that the least noisy data present this signature and that the r.m.s. is of order of $0.25\Delta$, which is 
consistent with the prediction of this approach.\cite{comment2}
Furthermore, it is clear from the figures that 
{\em the temperature effect is the main cause of the deviation from the CEI model result}.
\begin{figure}
\begin{center}
\leavevmode
\rotatebox{-0}{
\epsfxsize = 7cm 
\epsffile{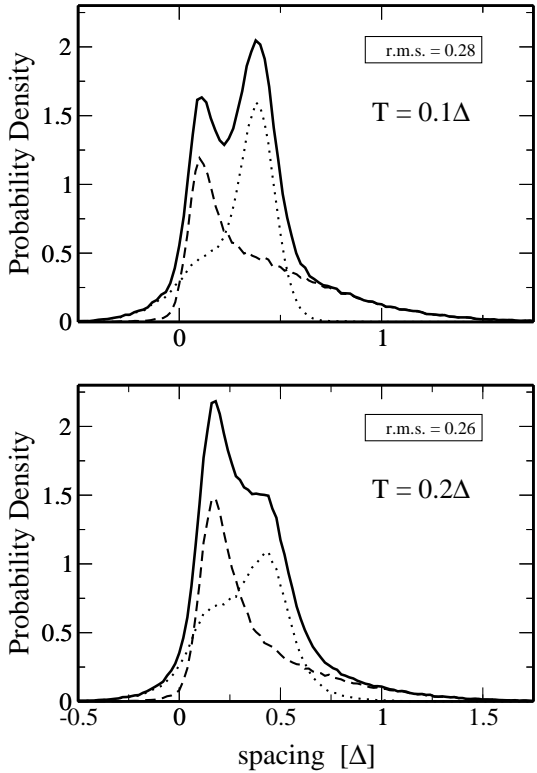}
}
\end{center}
\caption{Finite temperature CB peak spacing distribution obtained from the Hamiltonian (Eq.~\ref{leading}) with 
$N\smeq500$ ($g \!\approx\! 6$), $J_S \smeq 0.32 \Delta$ and $\kt \smeq 0.1 \Delta$ $(0.2 \Delta)$ in the top (bottom) plot. 
Notice that both the 
even/odd effect and the asymmetry are small.} 
\end{figure}

In conclusion, we have shown that the presence of the exchange interaction imposes a more restrictive condition on $T$ for 
observing GS properties in QDs. In particular, most of the experiments done so far require including temperature effects for 
their interpretation. The observed PSD seems to be the result of the addition of several small contributions.
 
We appreciate helpful discussions with D. Ullmo, L. I. Glazman and I. L. Aleiner.
GU acknowledges partial support from 
CONICET (Argentina). 
This work was supported in part by the NSF (DMR-0103003).


\begin{thebibliography}{10}

\bibitem{KouwenetalRev97}
L.~P. Kouwenhoven, C.~M. Marcus, P.~L. McEuen, S. Tarucha, R.~M. Westervelt,
  and N.~S. Wingreen,  in {\em Mesoscopic Electron Transport}, edited by L.~L.
  Sohn, L.~P. Kouwenhoven, and G. Sch{\"o}n (Kluwer, New York, 1997), pp.\
  105--214.

\bibitem{SivanBAPAB96}
U. Sivan, R. Berkovits, Y. Aloni, O. Prus, A. Auerbach, and G. Ben-Yoseph,
  Phys. Rev. Lett. {\bf 77},  1123  (1996).

\bibitem{PatelCSHMDHCG98}
S.~R. Patel, S.~M. Cronenwett, D.~R. Stewart, A.~G. Huibers, C.~M. Marcus,
  C.~I. Duruoz, J.~S. Harris, K. Campman, and A.~C. Gossard, Phys. Rev. Lett.
  {\bf 80},  4522  (1998).

\bibitem{PatelSMGASDH98}
S.~R. Patel, D.~R. Stewart, C.~M. Marcus, M. Gokcedag, Y. Alhassid, A.~D.
  Stone, C.~I. Duruos, and J. J.~S.~Harris, Phys. Rev. Lett. {\bf 81},  5900
  (1998).

\bibitem{SimmelAWKK99}
F. Simmel, D. Abusch-Magder, D.~A. Wharam, M.~A. Kastner, and J.~P. Kotthaus,
  Phys. Rev. B {\bf 59},  R10441  (1999).

\bibitem{LuscherHEWB01}
S. L{\"u}scher, T. Heinzel, K. Ensslin, W. Wegscheider, and M. Bichler, Phys.
  Rev. Lett. {\bf 86},  2118  (2001).

\bibitem{PrusAASB96}
O. Prus, A. Auerbach, Y. Aloni, U. Sivan, and R. Berkovits, Phys. Rev. B {\bf
  54},  R14289  (1996).

\bibitem{BlanterMM97}
Y.~M. Blanter, A.~D. Mirlin, and B.~A. Muzykantskii, Phys. Rev. Lett. {\bf 78},
   2449  (1997); Phys. Rev. B {\bf 63},  235315  (2001).

\bibitem{Berkovits98}
R. Berkovits, Phys. Rev. Lett. {\bf 81},  2128  (1998).

\bibitem{BrouwerOH99}
P.~W. Brouwer, Y. Oreg, and B.~I. Halperin, Phys. Rev. B {\bf 60},  R13977
  (1999).

\bibitem{BarangerUG00}
H.~U. Baranger, D. Ullmo, and L.~I. Glazman, Phys. Rev. B {\bf 61},  R2425
  (2000).

\bibitem{KurlandAA00}
I.~L. Kurland, I.~L. Aleiner, and B.~L. Altshuler, Phys. Rev. B {\bf 62},
  14886  (2000).

\bibitem{JacquodS00}
P. Jacquod and A.~D. Stone, Phys. Rev. Lett {\bf 84}, 4951 (2000); also cond-mat/0102100  (unpublished)

\bibitem{UllmoB01}
D. Ullmo and H.~U. Baranger, cond-mat/0103098 (unpublished).

\bibitem{VallejosLM98}
R.~O. Vallejos, C.~H. Lewenkopf, and E.~R. Mucciolo, Phys. Rev. Lett. {\bf 81},
   677  (1998); Phys. Rev. B {\bf 60}, 13682  (1999).

\bibitem{UsajB01_universal}
G. Usaj and H.~U. Baranger (unpublished).

\bibitem{OngBHPM01}
T.~T. Ong, H.~U. Baranger, D.~M. Higdon, S.~R. Patel, and C.~M. Marcus
  (unpublished).

\bibitem{AleinerBG01}
I.~L. Aleiner, P.~W. Brouwer, and L.~I. Glazman, cond-mat/0103008
  (unpublished) and references therein.

\bibitem{GlazmanM88}
L.~I. Glazman and K.~A. Matveev, Pis'ma Zh. \'Eksp. Teor. Fiz. {\bf 48},  403
  (1988), [JETP Lett. {\bf 48}, 445 (1988)].

\bibitem{Beenakker91}
C.~W.~J. Beenakker, Phys. Rev. B {\bf 44},  1646  (1991).

\bibitem{Akera99a}
H. Akera, Phys. Rev. B {\bf 59},  9802  (1999).

\bibitem{MeirWL91}
Y.~M. Meir, N.~S. Wingreen, and P.~A. Lee, Phys. Rev. Lett. {\bf 66},  3048
  (1991).

\bibitem{FolkMH00}
J.~A. Folk, C.~M. Marcus, and J.~S. Harris, cond-mat/0008052 (unpublished).

\bibitem{comment} 
This coefficient is difficult to evaluate since that requires calculating the electrostatic potential of a set of conductors 
(QD+gates) in a particular geometry. However, it is possible to estimate upper and lower bounds on its value.
\cite{UsajB01_universal}
For an {\em isolated} QD it can be evaluated explicitly, giving $b_{00}\!\simeq\!0.002$ for an ellipsoidal geometry. 

\bibitem{AlhassidM99}
Y. Alhassid and S. Malhotra, Phys. Rev. B {\bf 60},  R16316  (1999).

\bibitem{comment2}
In the experiments cited, $T$ is greater than $0.2\Delta$ but the experimental 
noise should also be included. 

\end{thebibliography}
\end{document}